\documentstyle[aps,preprint,12pt,prl]{revtex}
\begin{document}
\draft
\title{Does a magnetic field modify the critical behaviour at the
metal-insulator transition in 3-dimensional disordered systems?}
\author{E. Hofstetter}
\address{Blackett Laboratory, Imperial College, London SW7 2BZ, UK}
\author{M. Schreiber}
\address{Fachbereich Physik, Technische Universit\"{a}t
Chemnitz-Zwickau, PSF 964, \\
D-09009 Chemnitz, Federal Republic of Germany}
\date{\today}
\maketitle
\begin{abstract}
The critical behaviour of 3-dimensional disordered systems with magnetic
field is investigated by analyzing the spectral fluctuations of the energy
spectrum. We show that in the thermodynamic limit we have two different
regimes, one for the metallic side and one for the insulating side with
different level statistics. The third statistics which occurs only exactly
at the critical point is {\it independent} of the magnetic field. The
critical behaviour which is determined by the symmetry of the system {\it at}
the critical point should therefore be independent of the magnetic field.
\end{abstract}
\vspace{0.5cm}
\newpage
It is now well known that the introduction of disorder into periodic structures
has dramatic effects on the properties of the system. In particular 3
dimensional (3D) systems exhibit a metal-insulator transition (MIT) as a
function of the disorder. A lot of works \cite{MKK} have already been devoted
to the study of this phase transition, theoretically, numerically, as well as
experimentally. While all the methods seem to agree on the fact that the MIT
should only be characterized by the fundamental symmetries of the system and
that one can expect a second order phase transition, its description, e.g.
with respect to critical disorder and critical exponents, still remains
a controversial object of discussions \cite{MKK}.
\par
Theoretically the standard method applied, namely the $2+\epsilon$
expansion, for the calculation of the critical exponents turns out to be
inadequate for $d=3$ \cite{IL,FW}. It seems to be a characteristic of
several theories for disordered systems using perturbative methods, like
the nonlinear $\sigma$-model \cite{FW1} or the self-consistent theory
\cite{DV}, to provide quantitative results in the weak localization regime
(i.e. for small disorder) but to fail in the region of the critical disorder.
\par
Because of this problem numerical investigations in the frame of the Anderson
model of localization have played an important role in the description of the
MIT. Based on extensive computations by means of the transfer matrix method
(TMM) \cite{MK} it is now usually accepted that the critical exponent is
$\nu \simeq 1.4$ and is independent of the distribution chosen in the
Hamiltonian \cite{EH,MK1}. This result has been confirmed recently
\cite{EH1,BS} using a completely different method, namely the energy level
statistics method (ELSM). Meanwhile other TMM studies have been carried out
on similar models but with magnetic field, which changes the symmetry, driving
the system from the orthogonal to the unitary universality class, due to the
break of the time reversal invariance.
Surprisingly, in spite of this change of universality class, the same value of
the critical exponent has been found with and without magnetic field and that
independent of the strength of the magnetic field \cite{BK,BK1}. In this
letter,
using the ELSM which will again turn out to be very suitable for such a study,
we propose to explain this surprising observation with symmetry arguments
shedding new light on the problem. Our conclusions are based on our numerical
results for the spacing distribution $P(s)$ and the $\Delta_{3}$ statistics
showing that there is a critical enensemble (CE) charateristic for the
MIT irrespective of the magnetic field which implies that the critical 
exponent should be independent of the magnetic field.
\\
\par
In order to investigate the MIT with magnetic field we consider the usual
Anderson Hamiltonian with an additional phase factor in the off-diagonal
elements,
\begin{equation}
H=\sum_{n}\epsilon_{n} |n\rangle \langle n| +
\sum_{n\neq m} e^{i\theta_{n,m}} |n\rangle \langle m|,
\end{equation}
where the sites {\it n} are distributed regularly in 3D space, e.g. on a simple
cubic lattice, with periodic boundary conditions, and 
$\theta_{n,m}=-\theta_{m,n}=\theta$. Only interactions with 
the nearest neighbours are considered.
The site energy $\epsilon_{n}$ is described by a stochastic variable. In the
present investigation we use a box distribution with variance
$\sigma^{2}=W^{2}/12$. $W$ represents the disorder and is the critical
parameter. $\theta=0$ describes the case without magnetic field with a
Hamiltonian invariant under orthogonal transformation, while $\theta\neq 0$
corresponds to the magnetic field case which is invariant under unitary
transformation. In spite of the simplicity of the model, it contains all the
relevant properties necessary to describe the MIT with magnetic field. A similar
Hamiltonian has been proposed by Pandey {\it et al.} \cite{AP} and used to
study, by ELSM, the transition from the Gaussian orthogonal ensemble (GOE) to
the Gaussian unitary ensemble (GUE) in a metallic ring pierced by a magnetic
flux \cite{GM}.
\par
Based on this Hamiltonian, the MIT in presence of a magnetic field will be
studied by the ELSM, i.e. via the fluctuations of the energy spectrum. This
method has already given very interesting results in the case $\theta=0$,
where the MIT corresponds to a transition from the GOE to the Poisson ensemble
(PE) \cite{BA,BAL} which reflects completely uncorrelated energy levels in the
localized regime. In the thermodynamic limit one obtains two different
regimes: GOE for $W<W_{c}$ and PE for $W>W_{c}$, which are separated by a
critical ensemble (CE) \cite{BS,EH2} occuring at the critical disorder
$W_{c}$. For $\theta \neq 0$ one can expect a transition between GUE and PE,
but the crucial question is what happens in the vicinity of the critical
point.
\par
Before giving the results we shortly review the ELSM. Starting from Eq.(1)
the energy spectrum was computed by means of the Lanczos algorithm (which
is suited to diagonalize such very sparse secular matrices) for systems of
size $M \times M \times M$ with $M=13 \;{\rm and}\; 21$, disorder $W$ ranging
from 3 to 80 and phase $\theta=0.1\pi$. The number of different realizations
of the random site energies $\epsilon_{n}$ was chosen so that about
$2 \cdot 10^{5}$ eigenvalues were obtained for every pair of parameters
$(M,W)$ which means between 25 and 90 realizations, for which the full spectrum
has been computed. For the subsequent investigations only half of the spectrum
around the band center is considered so that the results are not deteriorated
by the strongly localized states near the band edges. After unfolding the
obtained spectrum the fluctuations can be appropriately characterized \cite{OB}
by means of the spacing distribution $P(s)$ and the Dyson-Metha statistics
$\Delta_{3}$. $P(s)$ measures the level repulsion, it is normalized, so is
its first moment because the spectrum is unfolded. $\Delta_{3}$ which measures
the spectral rigidity is given by
\begin{equation}
\Delta_{3}(L)=\left\langle \frac{1}{L} \min_{A,B} \int_{\varepsilon '}
^{\varepsilon '+L}(N(\varepsilon)-A\varepsilon-B)^{2}
\;d\varepsilon \right \rangle_{\varepsilon '}  .
\end{equation}
where $\langle \; \rangle_{\varepsilon '}$ means that we average over
different parts of the spectrum.
\par
Using the random matrix theory (RMT) it is possible to calculate $P(s)$ and
$\Delta_{3}$ for the two limiting cases of the spectrum, namely the GUE and
the PE. For the metallic side one obtains \cite{OB}
\begin{equation}
P_{GUE}(s) \simeq \frac{32}{\pi^{2}} s \exp(-\frac{4}{\pi} s^{2})
\end{equation}
\begin{equation}
\Delta_{3}(L)=\frac{2}{L^{4}} \int^{L}_{0} %
(L^{3}-2L^{2}r+r^{3})\Sigma_{GUE}^{2}(r) \;dr
\end{equation}
where $\Sigma_{GUE}^{2}(r)$ is the variance of the number of levels in a
spectral window of width $r$ and is given by
\begin{equation}
\Sigma_{GUE}^{2}(r)=\frac{2}{\pi^{2}}\left[\ln (2\pi r)+\gamma +1
-\cos(2\pi r)-{\rm Ci}(2\pi r)+
2r\left( 1-\frac{2}{\pi}{\rm Si}(2\pi r)\right) \right]
\end{equation}
The formula (3) for $P_{GUE}(s)$ is exact only in the case of $2 \times 2$
matrices but remains a good approximation for the other cases. For $\Delta_{3}$
there is no analytical solution and the integral (4) has to be calculated
numerically.
\par
For the localized case we have
\begin{equation}
P_{PE}(s)=e^{-s} ,
\end{equation}
\begin{equation}
\Delta_{3}(L)=\frac{L}{15}.
\end{equation}
\par
In Fig.\ref{1} the results for the Dyson-Metha statistics are reported. We 
find, as expected, the GUE and the PE regimes for small and large
disorder respectively as well as the continuous transition between them as a
function of $W$. In a previous work \cite{EH2} it was shown for the case
$\theta=0$ that the $\Delta_{3}$-curves are functions of the system size except
at the critical point where they are size independent. Moreover the curves
were shown to move with increasing $M$ towards the GOE for $W<W_{c}$ and
towards the PE for $W>W_{c}$. In Fig.\ref{2} we observe a similar behaviour
for $\theta=0.1\pi$, accordingly this time the curves move towards the
GUE and the PE. The critical value of the disorder where the curves are size
independent turns out to be $W_{c}\simeq 16.5$. Correspondingly, in the
thermodynamic limit we expect two different regimes with the GUE and the PE,
separated by the CE at the MIT. This has now to be compared to the CE obtained
in the case without magnetic field. In Fig.\ref{3} one sees that the CE curves
are identical and thus the symmetry should be the same, too.
Moreover this was checked for $\theta=\pi/2$ which is the "most Hermitian" case 
and the same results were obtained (cp. Fig.\ref{3}). Only the critical 
disorder, which is not a universal value, is slightly shifted upward towards 
$W_{c}\simeq16.65$. Such a behaviour has already been noted in a different 
numerical approach \cite{BK1}.
\par
The CE curve seems to be proportional to $L^{3/4}$ for small $L$ and becomes
linear when increasing $L$. This would mean that the shape of the curve cannot
solely be described neither by the results of Kravtzov {\it et al.} \cite{VK}
nor by the results of Alt'shuler {\it et al.} \cite{BAL} but rather would be
in agreement with Ref.\cite{VK} for small $L$ and with Ref.\cite{BAL} for large
$L$. Moreover it has to be stressed again that the CE curve is completely
independent of the magnetic field in contrast to Ref.\cite{VK}.
\par
\begin{figure}
\caption{Spacing distribution $P(s)$ for $M=21$, $W=3,80$ and $\theta=0.1\pi$.
The histograms are the numerical results and the full lines reflect the two
expected limiting ensembles, namely the GUE for the metallic side and the PE
for the insulating side. The points $\circ , \ast$ represent $P(s)$ at the
critical point for $M=21$ and $M=13$. The symbol $\diamond$ denotes the CE for
the case without magnetic field ($\theta=0$).}
\label{4}
\end{figure}
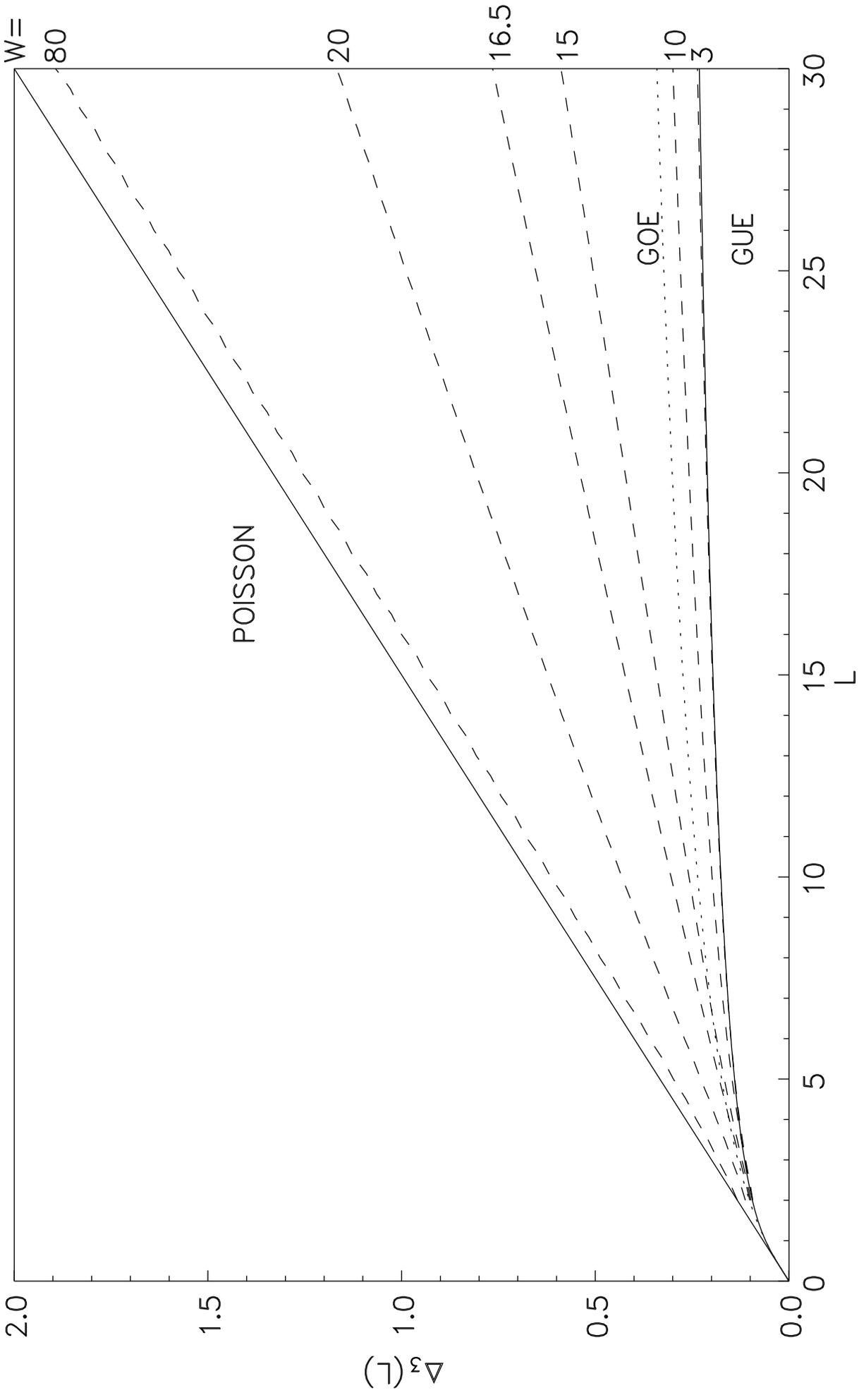
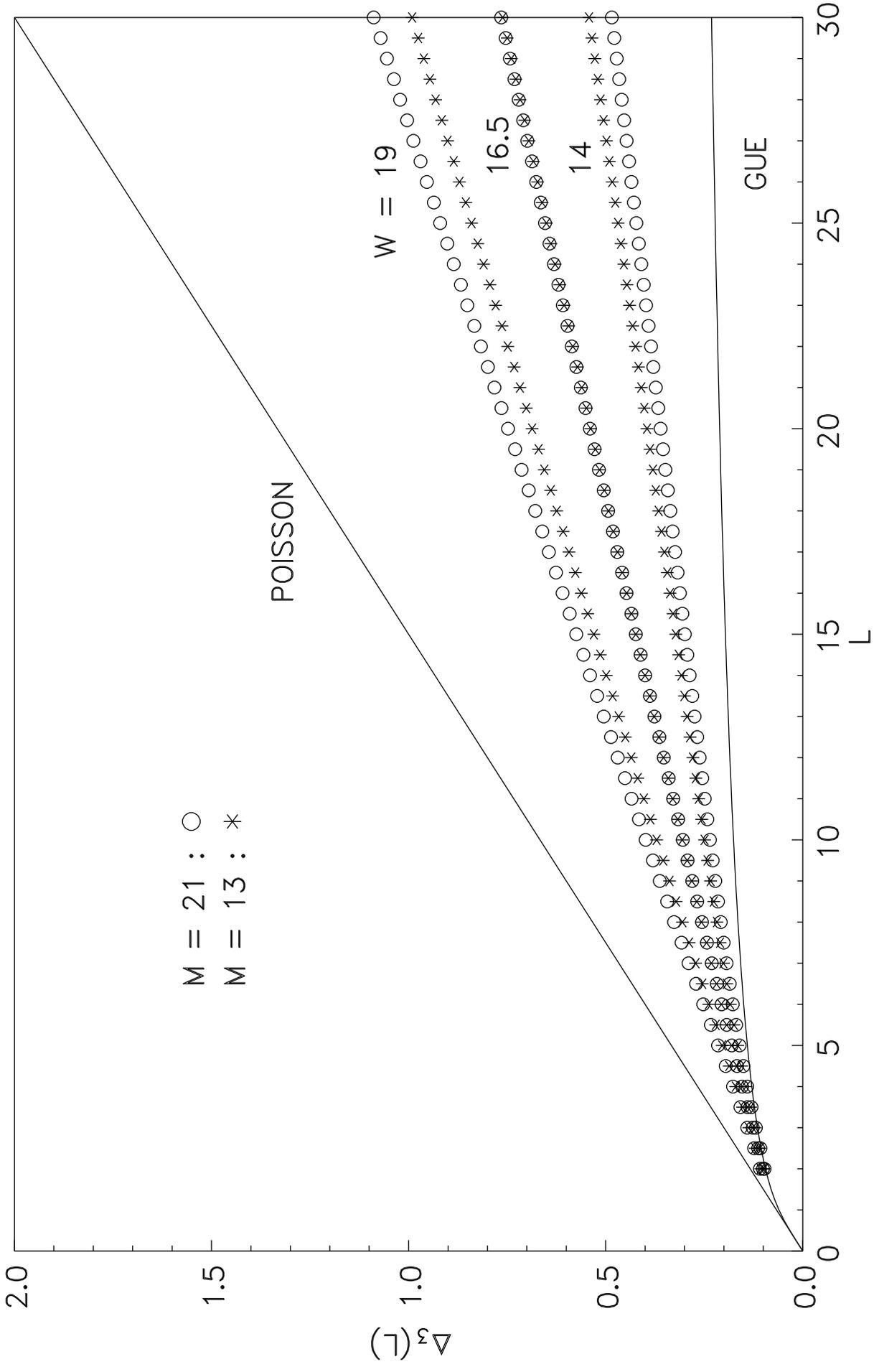
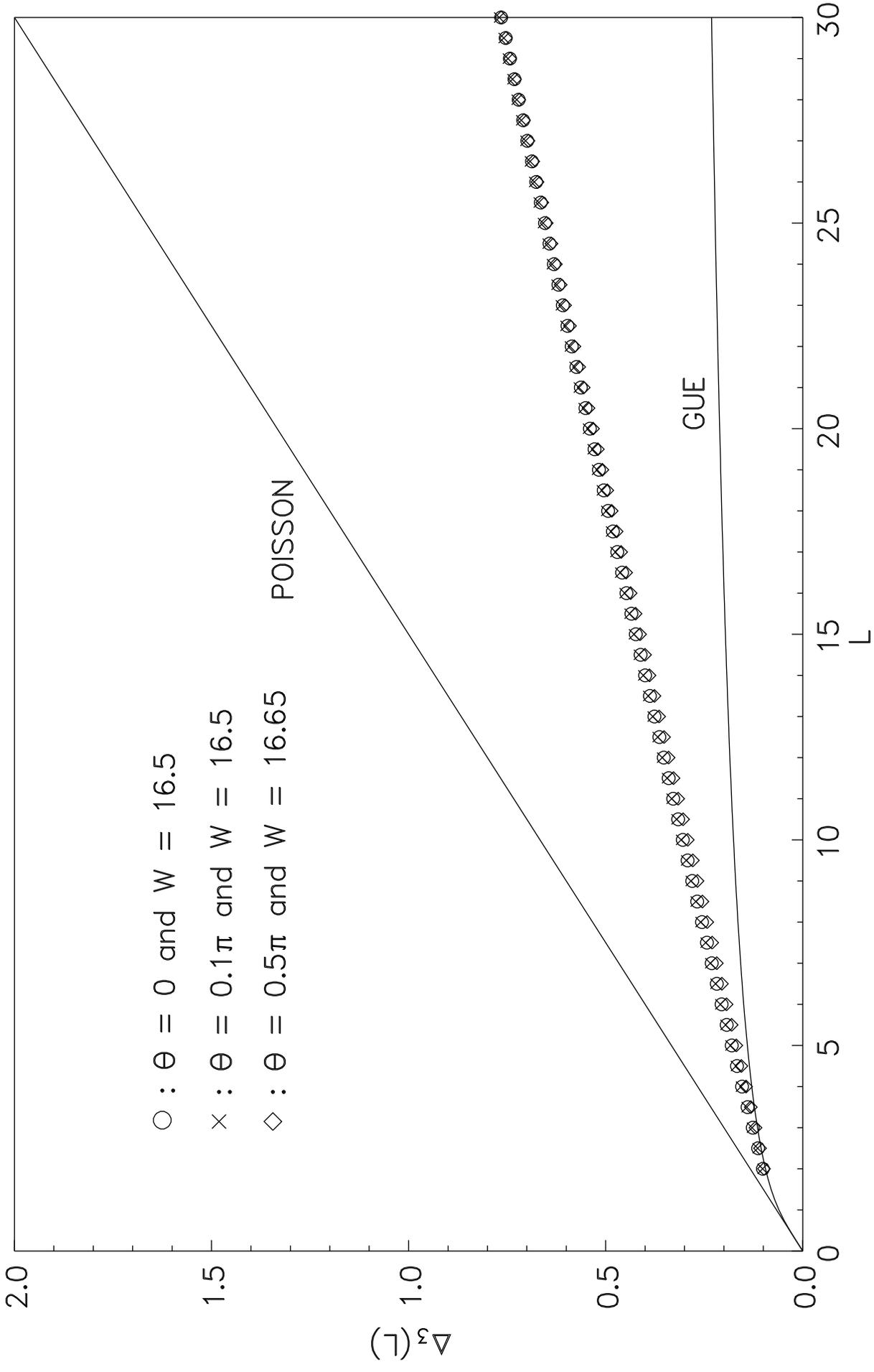
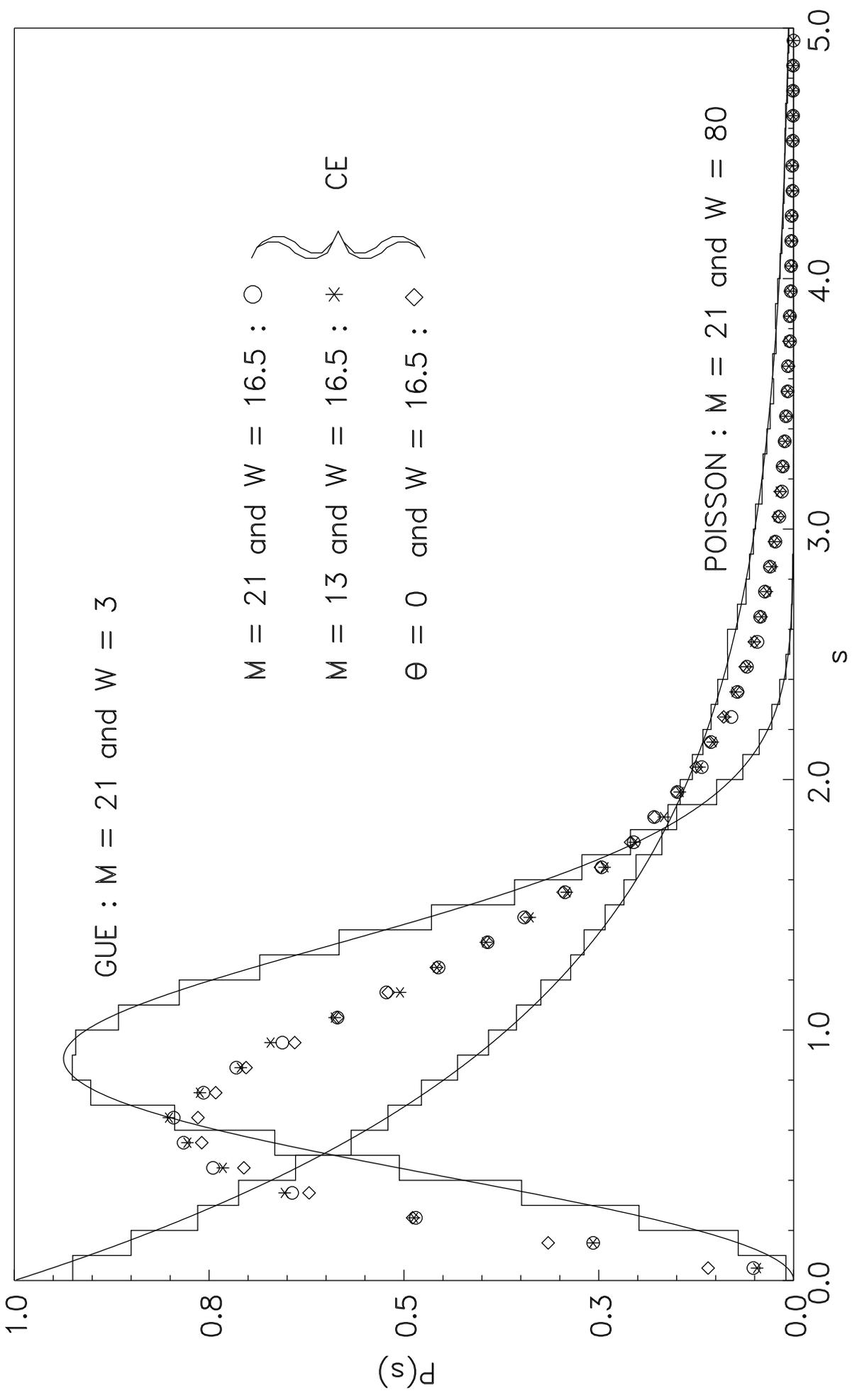

The same results are derived from the study of $P(s)$ in Fig.\ref{4}. For 
small and large disorder we again find the GUE and the PE respectively while
$P(s)$, at the critical point, is independent of $M$ and $\theta$.
Another interesting point is the controversy \cite{BS,AA} about the asymptotic 
behaviour of $P(s)$. A careful analysis of the shape of $P(s)$ at the critical 
point gives
\begin{equation}
P(s)=As\exp (-Bs^{\alpha})
\end{equation}
with $\alpha = 5/4 \pm 0.05$ and A,B the normalization constants which supports
the results obtained in Ref.\cite{AA}. Moreover one notes that the value of 
$\alpha$ is in very good agreement with the formula $\alpha=1+1/\nu d$ 
obtained by Kravtzov {\it et al.} \cite{VK} relating  $\alpha$ to the 
critical exponent $\nu$ and the dimension $d$ of the system. Finally it has
to be stressed that the dependence of the results in \cite{VK,AA} on $\beta$ 
($\beta=1$ for the GOE and 2 for the GUE) via some coefficients has 
not been derived from the calculation but
{\it assumed} from the very beginning leaving open the question whether those
results, at the critical point, depend on the magnetic field or not. Figure 
\ref{4} demonstrates that this is indeed the case.
\par
As the critical behaviour is determined by the fundamental symmetries of the
system one should consider the symmetry at the critical point. From the results
in Fig.\ref{3} and \ref{4} we conclude that the magnetic field has no influence 
on the critical behaviour. This means also that the critical exponent
$\nu=1.34\pm 0.10$, which was obtained \cite{EH1} for the $\theta=0$ case, is
transferable to the case with magnetic field.
\par
Finally we compare these results with experiments. Recent measurements performed 
on the persistent photoconductor ${\rm Al}_{0.3}{\rm Ga}_{0.7}{\rm As}$ 
\cite{SK} and on uncompensated Si:P \cite{HS} suggest, although the situation 
is still not completely clear, that the MIT is effectively  independent of 
the magnetic field. About the value of the critical exponent the 
situation is not really clearer but it can be mentioned that in Ref.\cite{HS}, 
considering carefully the range of the critical behaviour, a critical exponent 
of $\nu \simeq 1.3$ has been found which would be in agreement with our results.
This seems to indicate, at least in some cases, that the MIT is in fact driven 
purely by the disorder. Then it is not necessary to include interactions to
characterize the critical behaviour although these can be important outside
the critical regime. It has to be noted that some reserves have been made
\cite{HS1} about the results obtained in Ref.\cite{HS} and it will be 
interesting to check carefully this problem in order to clarify the situation.
\par
In conclusion we have shown, using the statistical properties of the energy
spectrum of the Anderson model of localization, that the MIT, which is
determined by the symmetry of the system {\it at} the critical point, is not
influenced by the magnetic field. This means that both cases, with or without
magnetic field belong to the same universality class, described by the CE,
which opens new perspectives in the comprehension of the MIT.
Although this universality class is defined only for the critical point it is
sufficient to fix the properties of the critical behaviour. This is in agreement
with previous numerical results \cite{BK,BK1} and recent experiments
\cite{SK,HS}. We point out that these results are valid in 3D systems and do
not apply to 2D where we know that the MIT is driven by the magnetic field
\cite{BK2}.

\newpage
\begin{figure}
\caption{Dyson-Metha statistics $\Delta_{3}(L)$ for $M=21$. A continuous
transition from GUE to Poisson statistics occurs as a function of the disorder
$W$ (denoted on the right-hand side). The dotted line shows the GOE result for
comparison.}
\label{1}
\end{figure}
\begin{figure}
\caption{Dyson-Metha statistics $\Delta_{3}(L)$ for system sizes
$M=13\;\;{\rm and}\;\;21$ and $W=14,16.5, \;\rm{and}\; 19$.}
\label{2}
\end{figure}
\begin{figure}
\caption{The Dyson-Metha statistics $\Delta_{3}(L)$ at the critical
point for different values of $\theta$.}
\label{3}
\end{figure}
\end{document}